\documentclass[usenatbib]{mn2e}
\usepackage{graphicx}
\usepackage{fixltx2e}

\begin{document}
\topmargin-1cm

\newcommand{\tq}{t_Q}
\newcommand{\mh}{M_h}
\newcommand{\fon}{f_{\rm on}}
\newcommand{\mbh}{M_{\rm BH}}
\newcommand{\beff}{\langle b\rangle}
\newcommand{\msun}{M_{\odot}}

\title[QSO halo masses]{Constraints on the correlation between QSO
luminosity and host halo mass from high-redshift quasar clustering}
\author[White et al.]{Martin White${}^{1}$,
Paul Martini${}^{2,3}$ and
J.D. Cohn${}^{4}$ \\
${}^1$Departments of Physics and Astronomy,
University of California, Berkeley, CA 94720 \\
${}^2$Department of Astronomy, The Ohio State University,
Columbus, OH 43210 \\
${}^3$Center for Cosmology and Astroparticle Physics,
The Ohio State University, Columbus, OH 43210 \\
${}^4$Space Sciences Laboratory, University of California, Berkeley 94720}

\date{\today}
\maketitle

\begin{abstract}
Recent measurements of high-redshift QSO clustering from the Sloan Digital Sky
Survey indicate that QSOs at $z\sim 4$ have a bias $\langle b\rangle\simeq 14$.
We find that this extremely high clustering amplitude, combined with the
corresponding space density, constrains the dispersion in the $L-\mh$ relation
to be less than 50\% at 99\% confidence for the most conservative case of a 
100\% duty cycle. This upper limit to the intrinsic dispersion provides as 
strong a constraint as current upper limits to the intrinsic dispersion in the 
local $\mbh-\sigma$ relation and the ratio of bolometric to Eddington 
luminosity of luminous QSOs. 
\end{abstract}


\section{Introduction}
\label{sec:introduction}

It has recently become accepted that quasar activity and black hole growth
are an integral part of galaxy evolution, however a detailed understanding
of what triggers quasar activity and how they are fueled still eludes us.
The leading contender for the identity of luminous, high redshift QSOs is
that they are black holes fed by by major mergers of gas-rich galaxies
\citep{Car90}.  Recent incarnations of such models
\citep{HaiLoe98,CavVit00,KauHae00,WyiLoe02,hopkins06}
provide a good description of many observed properties of the QSO population.

The situation is particularly interesting at high redshift, where the
population of supermassive black holes that powers the QSOs is
growing rapidly \citep[see e.g.][for a recent review]{Fan06}.  To
further understand this important phase of black hole and galaxy
evolution we would like to build a model in which QSO activity is tied
to the evolving cosmic web of dark matter halos.  The relationship
between QSOs and dark matter halos, their environments and duty
cycles, can be constrained via observations of their space density and
large-scale clustering \citep{ColKai89,HaiHui01,MarWei01}.  These
constraints become particularly sensitive if the QSOs inhabit the
rarest, most massive halos for which the spatial clustering
depends strongly on halo mass \citep{Kai84,BBKS86,EFWD88,ColKai89}.

At redshifts $z<3$ the advent of large optical surveys for QSOs has led to
firm constraints on the clustering as a function of luminosity and redshift
\citep{Cro05,Hen06,PorMagNor04,PorNor06,Mye07a,Mye07b,Mye07c,Ang08,Pad08}.
With the Sloan Digital Sky Survey (SDSS) we are now able to measure the
clustering of QSOs well even at $z>3$
\citep[][see also earlier work by \citealt{Kun97,Ste97}]{She07}.
Interestingly, the correlation length of the QSO population increases rapidly
with redshift, from $r_0=16.90\pm 1.73$ at $z\simeq 3$ to $r_0=24.30\pm 2.36$
at $z\simeq 4$.
\citet{She07} demonstrate that they can fit the observed clustering and space
density of $z\sim 4$ QSOs with the model of \citet{MarWei01} provided the
$z\sim 4$ QSOs are relatively long lived ($\tq\sim 160\,$Myr) and inhabit
halos more massive than about $5\times 10^{12}\,h^{-1}M_\odot$. For this
calculation \citet{She07} assume that there is a monotonic relationship
between instantaneous QSO luminosity and halo mass, with no scatter.
The actual relation between instantaneous QSO luminosity and host halo mass is
expected to include some scatter.  Scatter is expected in several of the
relationships linking the QSO luminosity and the host halo mass: the
relationship between the host halo mass and galaxy bulge, in the relationship
between galaxy bulge and black hole mass, in the relationship between black
hole mass and peak luminosity and in the relationship between peak and
instantaneous QSO luminosity.  

Here we demonstrate that the very high correlation length measured by
\citet{She07}, when combined with the rapid increase in bias for the most
massive halos, strongly constrains the scatter between instantaneous QSO
luminosity and halo mass.
The essential idea is that any scatter increases the contribution from lower
mass (and less highly biased) halos, so that a measurement of large clustering
amplitude limits the contribution from lower mass objects.
This constraint is as strong as direct, observational constraints at lower
redshift on the amount of scatter in the various relationships mentioned above.

We shall work throughout in the $\Lambda$CDM framework and adopt the following
cosmological parameters:
$\Omega_{\rm mat}=0.25$, $\Omega_\Lambda=0.75$, $h=0.72$ and $\sigma_8=0.8$.
Where appropriate we shall comment on how our results depend on these
particular choices.  The next section outlines our formalism and applies it
to the data of \citet{She07}, while \S\ref{sec:discussion} interprets our
results and discusses future observations.

\section{Formalism and Application to High-Redshift QSOs}  
\label{sec:formalism} 

\subsection{The clustering and abundance of QSOs}

\begin{figure}
\begin{center}
\resizebox{3.5in}{!}{\includegraphics{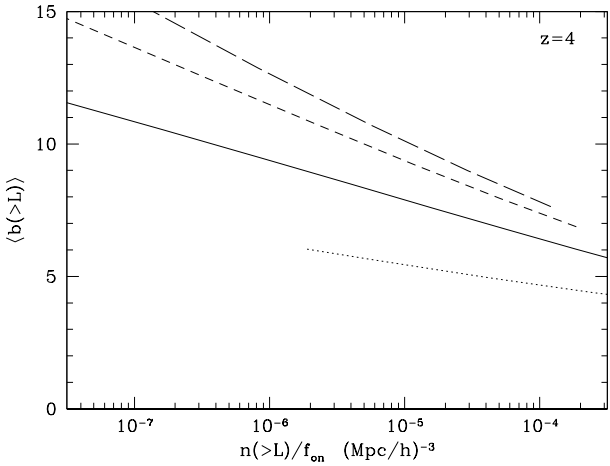}}
\end{center}
\vspace{-0.1in}
\caption{The large-scale bias, $\langle b\rangle$, vs.~the space density for
all QSOs at $z=4$ above a luminosity threshold assuming a log-normal
relationship between instantaneous QSO luminosity and halo mass with
dispersion $\sigma=0.1$ (long dashed), 0.5 (dashed), 1 (solid) and 2 (dotted).
The case $\sigma=0$ (not shown) is almost indistinguishable from $\sigma=0.1$.
In each case the lines run from $M_t=10^{12}$ to $10^{14.5}\,h^{-1}M_\odot$.}
\label{fig:b_vs_n}
\end{figure}

We will focus our attention on the space density and large-scale bias
of QSOs, as these are the easiest to interpret, observationally measurable 
properties of the population.  If the mean number of QSOs in a halo of 
mass $M_h$ is $N(M_h)$ then their number density and large-scale bias are
\begin{eqnarray}
\bar{n} &=& \int dM_h\ \frac{dn_h}{dM_h} N(M_h)\\
\langle b\rangle &=& \bar{n}^{-1}
  \int dM_h\ \frac{dn_h}{dM_h}b_h(M_h) N(M_h)
\label{eqn:nbcal}
\end{eqnarray}
where $dn_h/dM_h$ is the (comoving) number density of halos per mass interval
and $b_h(M_h)$ is the bias associated with halos of that mass.
For $dn_h/dM_h$ we use the fitting function from \citet{SheTor}.
The large-scale bias is slightly more problematic, as our results depend
upon massive halos being biased and different fits to $b_h(M_h)$ from
simulations have appeared in the literature.
\citet{SheTor} derived the bias appropriate to their mass function fit using
the peak background split.  A slightly higher bias at fixed mass comes from
assuming ellipsoidal collapse \citep{SheMoTor}.  The largest bias for halos
of a given mass is given by the bias of the \citet{PreSch} mass function,
as computed by \citet{ColKai89}, which is very similar to the fit of
\citet{Jin98} for the masses of interest.
These two forms give biases 30\% larger, in the mass and redshift range of
interest, than the lowest fit \citep[of][]{SheTor}. The fit of
\citet{Tin05} lies between that of \citet{SheMoTor} and \citet{SheTor}.  
For mass-thresholded samples of dark matter halos in a large N-body
simulation\footnote{The simulation evolved $1024^3$ particles in a
$1\,h^{-1}$Gpc box using the {\sl TreePM\/} code described in \citet{TreePM}.
See \citet{CodeCompare} for a recent comparison with other N-body codes.},
we found the \citet{PreSch} and \citet{SheTor} forms bracketed the
halo correlation function 
so this should represent the level of current uncertainty.
As it provides the most conservative estimate of the maximum allowed scatter,
and the numerical data lay closer to the form of \citet{PreSch}, we adopt this
as our fiducial bias.  However, we caution that the statistics in the
simulation are poor and we also illustrate the effect of choosing the
\citet{SheTor} form below.
In several recent simulations at other redshifts and masses the \citet{PreSch}
form appears to be a worse fit \citep{SheTor,SheMoTor,SelWar,Tin05}, but
unfortunately there are few direct N-body calibrations of $b_h(M)$ at the
number densities and redshifts of interest to us.  One could imagine that for
extremely rare peaks, whose collapse is much closer to spherical \citep{BBKS86},
these newer fits actually perform worse than the original formulation
\citep[see e.g.][]{CohWhi08,Dal08}.
This challenging numerical problem deserves further investigation.  

To interpret the large-scale clustering measurements of \citet{She07}
we then must specify $N(M_h)$.  We wish to choose as simple a
model as possible to illustrate the effect of scatter on the $\beff-\bar{n}$
relation, so we assume that the probability that a QSO is seen with
instantaneous luminosity $L$ is log-normally distributed\footnote{We use
natural logarithms throughout so that $\sigma$ can be interpreted as a
fractional scatter.  The log-base-10 based $\sigma$ would be
$\ln(10)\simeq 2.3$ times smaller.} around a central value $L_0(M_h)$ with
a width $\sigma_L$
\begin{equation}
  P\left(L|M_h\right) d\ln L = \frac{1}{\sqrt{2\pi}\sigma_L}
  \ \exp\left[ -\frac{\ln^2 \left(L/L_0\right)}{2\sigma_L^2} \right]
  \ d\ln L
  \qquad .
\end{equation}
If a given halo hosts a QSO, the probability that it is above some limiting
luminosity, $L_{\rm min}$, is then the integral of this expression from
$L_{\rm min}$ to infinity, which can be expressed as an error function.
If we assume that at most a fraction $f_{\rm on}$ of halos host active QSOs
at any epoch and that $L_0\propto M_h^\alpha$ in the region of the turn-on,
the mean number of QSOs above $L_{\rm min}$ in halos of mass $M_h$ is
\begin{equation}
  N(M_h) = \frac{f_{\rm on}}{2}{\rm erfc}\,
  \left[\frac{\ln M_t/M_h}{\sqrt{2}\sigma}\right]
\label{eqn:nofm}
\end{equation}
where $L_0(M_t)\equiv L_{\rm min}$, $\sigma=\sigma_L/\alpha$ and we
expect $\alpha\approx 1$.  We shall assume Eq.~(\ref{eqn:nofm}) in what
follows and try to constrain $\sigma$.
Figure \ref{fig:b_vs_n} shows $\langle b(>L)\rangle$
vs.~$\bar{n}(>L)/f_{\rm on}$ for a range of $M_t$ and $\sigma$.
The larger the dispersion the lower the bias at fixed space density,
as a larger fraction of the QSOs are hosted in lower mass halos.
Similarly, a decrease in $f_{\rm on}$ must be compensated by a decrease in
$M_t$ to hold $\bar{n}$ fixed, again leading to a lower $\beff$.
In our assumed cosmology the Universe is $t_U\simeq 1.6\,$Gyr old at
$z=4$.  If we assume $L_{\rm bol}/L_{\rm edd}=0.25$ \citep[e.g.][]{Kol06}
then the e-folding time is $\simeq 0.1\,t_U$, so
$f_{\rm on}\sim\mathcal{O}(1)$ is not unexpected.
While the inferred lifetimes are then weakly in conflict with the upper
range of observational constraints \citep[$10^6-10^8$ years;][]{Martini04}, 
these constraints largely stem from lower-redshift QSOs. 

Our model is not the most general one describing how QSOs inhabit dark
matter halos.  In particular, there is no reason in principle why halos
could not host more than one QSO.
To constrain the functional form of $N(M_h)$ would require modeling both
the large- and small-scale clustering, and additional assumptions about
the statistics of QSOs in halos.  This is beyond the scope of this paper,
so we merely note that the mass function is very steeply falling in the
range of halo masses that are of interest.  Our statistics are thus
dominated by the lowest mass halos in our sample.
In particular, the number density is approximately linear in $N(M_t)$.
Simply allowing $f_{\rm on}>1$ has no impact on $\langle b\rangle$, but
it would raise $\bar{n}$ and slightly weaken our constraints.  However
to increase $\bar{n}$ significantly above that for $f_{\rm on}=1$ would
require almost every halo of mass $M_t$ to host more than one high luminosity
QSO which would lead to an unacceptably high close pair fraction.
For example, if every halo hosted 2 QSOs the volume averaged correlation
function within the halo radius would be $\bar{\xi}\sim n_h^{-1} r_h^{-3}$.
Given the extreme rarity of these halos, this would be in conflict with
observations \citep[][see e.g.~Fig.~17 in \citealt{Hop08}]{Hen06,Mye07b}.

At this point it is also easy to see how a change in cosmological parameters
affects $\beff-\bar{n}$.  For example, if we raise $\sigma_8$, halos of a
fixed mass correspond to less rare peaks.  These halos are therefore slightly 
less biased and significantly more numerous.  To hold $\bar{n}$ fixed we
would need to increase $M_t$.  To hold the bias fixed requires that we increase 
$M_t$ yet further and therefore to match both the $\beff$ and $\bar{n}$ 
the duty cycle must increase.
For $\bar{n}\simeq 10^{-7}h^3{\rm Mpc}^{-3}$, a 10\% increase in $\sigma_8$
leads to a 4\% decrease in $\beff$, which is insignificant for our purposes.
In general, for a power-law power spectrum with index $n_{\rm eff}$, the
fractional change in number density at fixed bias is $6/(3+n_{\rm eff})$
times the fractional change in $\sigma_8$.  Small changes in the other
cosmological parameters have even smaller effects.  

\subsection{Comparison with observations}

This formalism allows us to 
use the observations of high-$z$ QSO clustering from \citet{She07}
to constrain $\sigma$ and $f_{\rm on}$ in our model.  To do so we need to
estimate the large-scale bias, $\beff$, from the measurements provided in
\citet{She07} and this requires making several assumptions.

\begin{figure}
\begin{center}
\resizebox{3.5in}{!}{\includegraphics{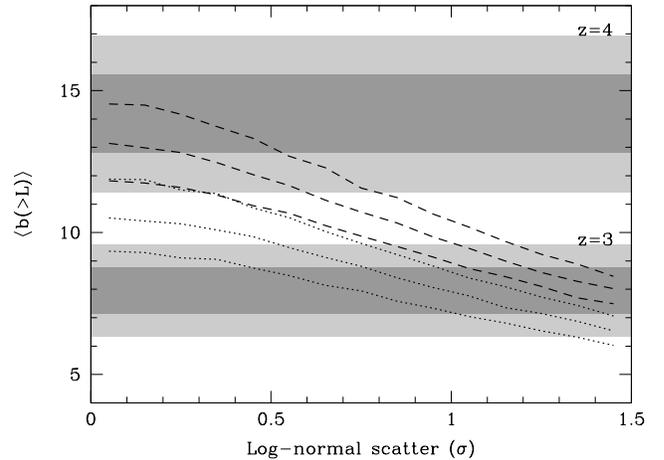}}
\end{center}
\vspace{-0.1in}
\caption{The large-scale bias, $\langle b\rangle$, vs.~log-normal dispersion,
$\sigma$, for all QSOs brighter than a given $L_{\rm min}$ chosen such that
$\bar{n}=5.6\times 10^{-7}\,h^3{\rm Mpc}^{-3}$ (lower, dotted lines) or
$2\times 10^{-7}\,h^3{\rm Mpc}^{-3}$ (upper, dashed lines).  The shaded bands
indicate the $\pm 1 ~ {\rm and } ~ 2\sigma$  range of bias measured by 
\protect\citet{She07}, as described in the text.  For each triple of lines
the upper line assumes $f_{\rm on}$ of Eq.~(\protect\ref{eqn:nofm}) is unity
while the lower lines assume $f_{\rm on}=0.3$ and $0.1$.}
\label{fig:b_vs_sig}
\end{figure}

First we must convert from the clustering quoted by \citet{She07} to
large-scale bias.  There are several ways to do this, and we have chosen to
use the amplitude of their fits to $r^{-2}$ power-laws over the range
$5<r<20\,h^{-1}$Mpc.  At lower $z$ the bias is relatively constant over this
range and the slope of the dark matter correlation function is close to $-2$,
so converting the fit into a measurement of $\langle b\rangle$ is
straightforward.  However at $z>3$ the QSOs are hosted by increasingly
rare halos, whose bias is becoming more scale dependent.  Using the mass
and halo catalogs from the N-body simulation described previously we find that
the mass correlation function is shallower than $r^{-2}$ below
$\mathcal{O}(10\,h^{-1}{\rm Mpc})$ but the halo correlation function is still
quite close to $r^{-2}$ down to $5\,h^{-1}$Mpc.
We therefore convert from $r_0$ to $\langle b\rangle$ by matching
\begin{equation}
  \xi_{qq}=\left(\frac{r_0}{r}\right)^2=\langle b\rangle^2\xi_{\rm dm}(r)
\end{equation}
at $r=20\,h^{-1}$Mpc, where the bias is close to constant.
(This is also the scale where we compared the simulations to the \citet{PreSch}
fitting form to obtain the bias.)  
For convenience we use the non-linear power spectrum of \citet{HaloFit} when
computing $\xi_{\rm dm}$, but we checked that this agrees well with the results
of the N-body simulation on the scales of interest and the correction for
non-linearity is relatively small.

The clustering measurements of \citet{She07} are averaged over bins in 
redshift, over which the bias and mass correlations are evolving strongly.
Fortuitously, the effects approximately cancel, leading to a slow evolution
in $\xi_{qq}$.  We estimate the effects of clustering evolution on our
constant-time constraints by considering 3 models for the evolution:
passive evolution (which for massive, rare halos corresponds almost to
$\xi_{qq}=$constant), constant bias and constant halo mass.  Each model
predicts $\xi_{qq}(r,z)$ from which we can compute the average value
measured by \citet{She07} as
\begin{equation}
  \langle\xi(r)\rangle \equiv
  \frac{\int dz\ (dN/dz)^2(H/\chi^2)\xi(r,z)}{\int dz\ (dN/dz)^2(H/\chi^2)}
\end{equation}
where $dN/dz$ is the redshift distribution, $H$ is the Hubble parameter at
redshift $z$ and $\chi$ is the comoving angular diameter distance to
redshift $z$.  Using $dN/dz$ from Table 1 of \citet{She07} in the two bins
$2.9<z<3.5$ and $z>3.5$ we find that for the passive evolution model the
correlation length inferred from $\langle\xi\rangle$ is within 1\% of that
inferred from $\xi(z=3)$ and $\xi(z=4)$ respectively.
For the constant halo mass case $\xi(z)$ increases with $z$, but the inferred
$r_0$ is still within 5\% of the constant $z$ value for both samples.
For the constant bias case, $\xi(z)$ decreases with $z$, but the inferred
$r_0$ is within 4\% of the constant $z$ value for both samples.
Since the quoted errors on $r_0$ are larger than this, we shall interpret the
quoted correlation lengths as measurements at $z=3$ and $z=4$ respectively.

Assuming $\xi_{qq}=(r_0/r)^2$, \citet{She07} quote
$r_0=(16.90\pm 1.73)\,h^{-1}$Mpc for their ``good'' sample with $2.9<z<3.5$.
Using the dark matter correlation function for our cosmology at $z=3$ this
corresponds to $\langle b\rangle=7.9\pm 0.8$.
The quoted space density is $\bar{n}=5.6\times 10^{-7}\,h^3{\rm Mpc}^{-3}$.
If we use the `all' sample both $r_0$ and hence $\langle b\rangle$ are 14\%
lower.  We will find that this point is relatively unconstraining for either
choice.

At $z=4$ the best fit is $r_0=(24.30\pm 2.36)\,h^{-1}$Mpc, again using the
``good'' sample, corresponding to $\langle b\rangle=14.2\pm 1.4$, and
$\bar{n}=10^{-7}\,h^3{\rm Mpc}^{-3}$. 
The other clustering fits provided by \citet{She07} are discussed below.
Note however, as pointed out by \citet{She07}, this space density is actually
an underestimate of the true space density because the \citet{Ric06}
parameterization of the LF underestimates the measured space density from
$z\sim 3\rightarrow 4$ \citep[see Figure 20 of][]{Ric06}.  At $z\simeq 4$ this
underestimate is approximately a factor of two and we therefore use
$\bar{n}=2\times 10^{-7}\,h^3{\rm Mpc}^{-3}$.  The higher space density leads
to a smaller predicted $\langle b\rangle$, requiring smaller dispersion
$\sigma$ between luminosity and halo mass to fit the observed clustering
(i.e.~this provides a more stringent constraint).  

In Fig.~\ref{fig:b_vs_sig} we show $\langle b(>L)\rangle$ as a function of
log-normal scatter, $\sigma$, at fixed $\bar{n}$ for $f_{\rm on}=1$, 0.3 and
0.1.  The $68\%$ and $95\%$ confidence ranges in $\beff$ (above) are also shown.
As can be seen,
in order to get the large $\beff$ seen by \citet{She07}, the scatter at
$z\simeq 4$ has to be less than 80\% and the fraction of halos containing
quasars, $f_{\rm on}$, must be larger than $\sim10\%$.

The constraints will change if different fits to the clustering measurements,
different fits for $b_h$, or different number densities are used.
As the constraints are coming from $z=4$, we focus on this case.
\citet{She07} gives 4 fits to power-law correlation functions, for ``all''
or ``good'' QSO's, with fixed power law $(r_0/r)^2$ or varying power law.
The fiducial calculation above was for the ``good'' QSOs sample fit to
$(r_0/r)^2$, which has a $\chi^2/{\rm dof}=0.32$.  The other combinations
give: (sample, power-law index, $\chi^2$, $\delta b/b$)=
(``all'',  $2.00$, 0.52, -15\%),
(``good'', $2.14$, 0.32,  +6\%),
(``all'',  $2.28$, 0.50,  -6\%).
The central value in Fig.~\ref{fig:b_vs_sig} then just moves up and down by
the shift in bias (the error bars do change slightly in width). 
Our example in Fig.~\ref{fig:b_vs_sig} is one of the best $\chi^2$ cases but
also gives a high bias (and thus is very constraining).

\begin{figure}
\begin{center}
\resizebox{3.5in}{!}{\includegraphics{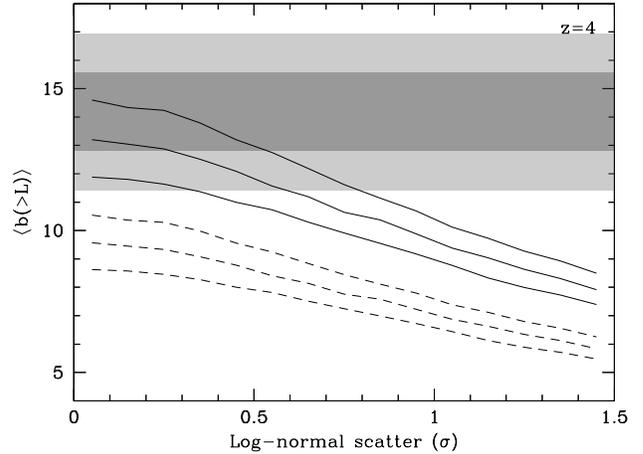}}
\end{center}
\vspace{-0.1in}
\caption{The same range of measured bias for $z=4$ as shown in
Fig.~\protect\ref{fig:b_vs_sig}, but now for two different analytic fits to
the theoretical bias.  The Press-Schechter (top, solid) bias fit is as before, 
the Sheth-Tormen (bottom, dashed) is shown below.
Again the shaded bands indicate the $\pm1$ and $2\sigma$ range of bias for
$z=4$ measured by \protect\citet{She07}, as described in the text.
For each triple of lines the upper line assumes $f_{\rm on}$ of
Eq.~(\protect\ref{eqn:nofm}) is unity while the lower lines assume
$f_{\rm on}=0.3$ and $0.1$.  If the Sheth-Tormen bias is used, there is no
overlap within the $2\sigma$ range of the clustering measurement.}
\label{fig:bias_cmp}
\end{figure}

As mentioned previously, our fiducial $b_h$ was chosen to give the least
stringent constraint.  The functional forms of \citet{SheTor,SheMoTor,SelWar},
and \citet{Tin05} predict lower $b_h$ at fixed $M_h$ and hence tighter
constraints on $\sigma$ and $f_{\rm on}$.
We illustrate the range in Figure \ref{fig:bias_cmp}, which compares the
results using our fiducial bias to that of \citet{SheTor}.  As we discussed
before the fits to $b_h(M_h)$ from N-body simulations, where they are
available, tend to lie between those of \citet{SheTor} and \citet{SheMoTor}
which differ from each other by $5-10\%$ in the mass and redshift range of
interest \citep[see also][]{SelWar,Tin05}.
Neither the Sheth-Tormen or Sheth-Mo-Tormen biases are able to fit the
measured bias and number density.

The most unconstraining estimate would be to use the lower amplitude of
clustering, e.g.~the ``all'' sample, at $z=4$.  We did not choose this because
there are a number of systematics which could lower the measured clustering
amplitude and the ``all'' sample has a worse $\chi^2$.  But if taken in
conjunction with the (most conservative) Press-Schechter bias and the highest
allowed number density, the limit can be weakened to allow $\sigma=1$ at 95\%
confidence for $f_{\rm on}=0.1$.  
It is also possible that QSOs may inhabit a special subclass of halos for which
the bias is larger than the average for that mass.
For instance there are indications that halo history affects clustering in
some cases \citep{Wec06,Wet07,GaoWhi07,CroGaoWhi07,JinSutMo07}.  
For the very rare and highly biased halos hosting QSOs at $z\simeq 4$ this
effect may be large enough to weaken our constraint. 
Unfortunately the bias of the relevant objects has not yet been measured in 
simulations.

\section{Discussion} \label{sec:discussion}

It appears that the most luminous, highest redshift QSOs have instantaneous
luminosities which are well correlated with their host halo masses.
The already strong constraint shown here should be improved by measurements 
of high-redshift QSO clustering with yet larger samples. New observations will 
broadly improve the constraint in two ways: through wider area observations 
to identify a larger number of QSOs similar in redshift 
to the SDSS sample, and through 
deeper observations over the same area. Deeper observations will identify 
slightly fainter QSOs at all redshifts and may provide a sufficient sample 
to measure the clustering amplitude at even higher redshift than $z \sim 4$, 
where we expect the bias to be even larger. Going wider in area will decrease 
the fractional error on $\beff$ for high-redshift QSOs. As these measurements 
of high-redshift QSO bias will remain Poisson limited, the fractional error on 
$\beff$ will scale as $1/N_{\rm qso}$ for a survey of $N_{\rm qso}$ QSOs. 
Going deeper to measure luminous QSOs at yet higher redshifts is also expected 
to be useful, as the data indicate strong evolution in the clustering 
amplitude with redshift. 

\begin{figure}
\begin{center}
\resizebox{3.5in}{!}{\includegraphics{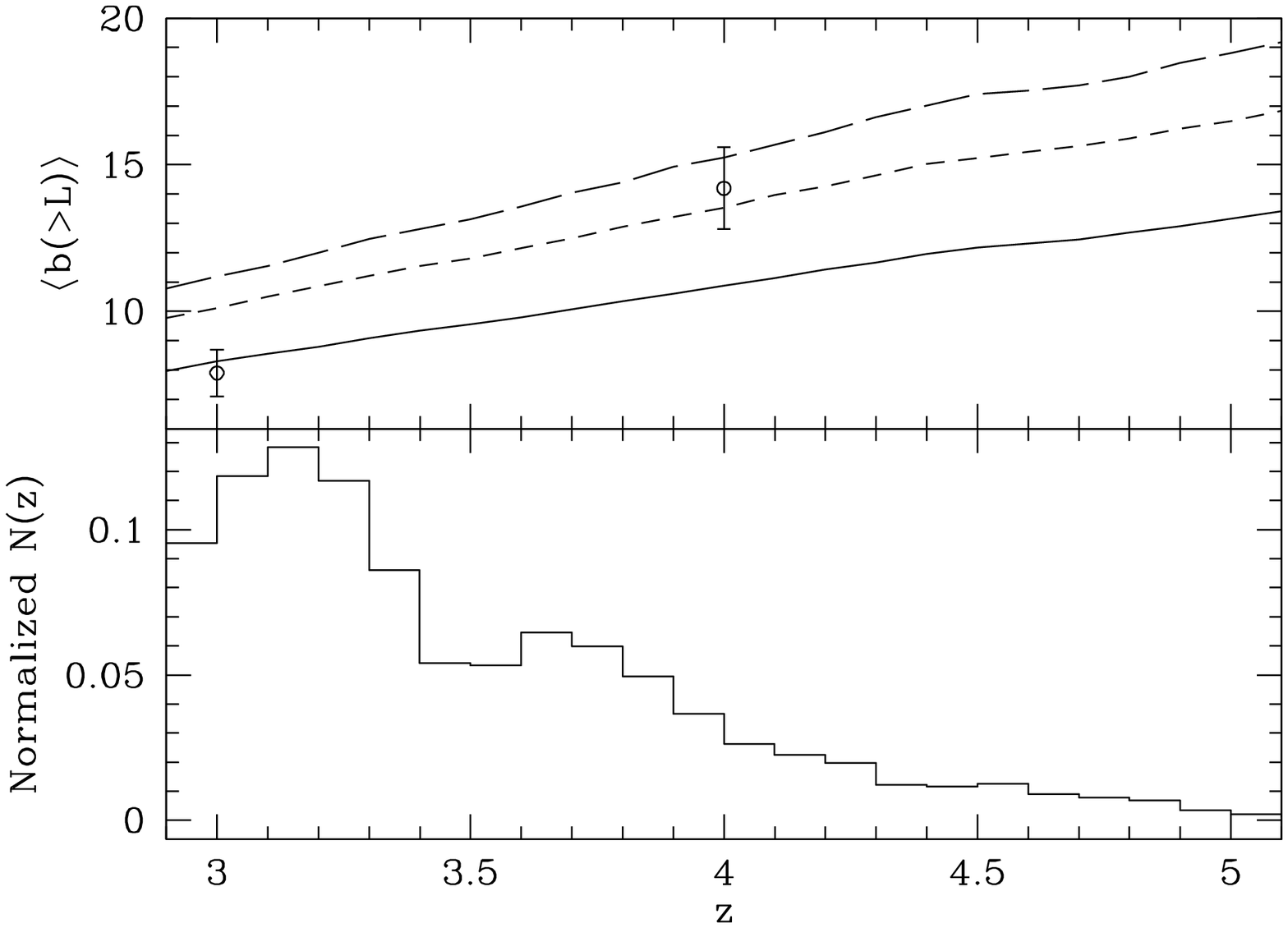}}
\end{center}
\vspace{-0.1in}
\caption{({\it Top}) The large-scale bias, $\langle b\rangle$, vs.~redshift 
of halos that match the integrated QSO space density from \protect\citet{Ric06} 
for $\fon = 1$ and dispersion $\sigma=0.1$ (long dashed), 0.5 (dashed), 
1 (solid) as in Figure~\ref{fig:b_vs_n}. Our bias estimates derived from 
the \protect\citep{She07} estimates are also shown. 
({\it Bottom}) Normalized number of QSOs per redshift bin from
\protect\citet{She07}. 
The rapid decrease in SDSS QSOs with redshift indicates that purely 
flux-limited surveys at $z > 3$ can not place as strong a constraint on 
$\sigma$ at yet higher redshifts, even though those QSOs are more strongly 
biased.}
\label{fig:b_vs_z}
\end{figure}

Figure~\ref{fig:b_vs_z} illustrates how $\beff$ increases with redshift for
three values of the scatter between QSO luminosity and host halo mass assuming
$\fon = 1$. At each redshift $\beff$ was calculated to match the QSO space
density evolution parametrized by \citet{Ric06}, although converted to our
cosmological parameters\footnote{For simplicity we did not correct the
\citet{Ric06} LF parameterization to match their binned LF here.  The space
density has therefore been slightly underestimated from $z=3\rightarrow 4$
and consequently the predicted bias is slightly higher.}.
The steepest relation between $\beff$ and redshift is for the case of the
smallest scatter. If $\fon = 1$ and the scatter is minimal, the relative value
of $\beff$ for different values of the dispersion is a good measure of the
fractional uncertainty in $\beff$ required to improve on our constraint.
These ratios demonstrate that the most effective constraint from clustering
arises from the lowest redshift at which $\fon = 1$ and minimal scatter is a
reasonable approximation. For example, the ratio of $\beff(\sigma=0.1)$ to
$\beff(\sigma=1.0)$ increases by less than 10\% from $z=4$ to $z=5$, while
the $N_{\rm qso}$ above the SDSS flux limit drops by approximately an order of
magnitude.  Therefore while the clustering amplitude increases, the decline in
$N_{\rm qso}$ for a flux-limited, high-redshift survey more than offsets this
gain.  By contrast, color selection of just very high-redshift (e.g.~$z>5$)
QSOs could produce a substantial improvement for a given total number of
spectra.

The immediate prospect for improvement in this constraint is completion of
the SDSS observations beyond the 4000 square degrees employed by \citet{She07}.
These observations will simply increase $N_{\rm qso}$ and the error bars should
scale as $N_{\rm qso}^{-1}$.
On a somewhat longer timescale, the proposed Baryon Oscillation Spectroscopic
Survey (BOSS\footnote{http://cosmology.lbl.gov/BOSS}) should approximately
double the number of $z>3.5$ QSOs, although it will mostly achieve this by
going deeper. Provided $\fon = 1$ and small scatter are still reasonable
approximations to this slightly fainter population (that will still be well
above the break in the QSO LF), BOSS' factor of two improvement in sample size
should decrease the error bars by approximately a factor of two and produce a
powerful improvement in the constraint.  For example, if the measured
bias stayed fixed and the errorbars decreased by a factor of two, the
upper limit on the observed scatter would improve to less than 0.5
(i.e.~0.2 dex) for $\fon=1$.
There are also upcoming photometric surveys that go even wider in area and
deeper: Pan-STARRS\footnote{http://pan-starrs.ifa.hawaii.edu}, 
the Dark Energy Survey (DES\footnote{http://www.darkenergysurvey.org}), and 
the Large Synoptic Survey Telescope (LSST\footnote{http://www.lsst.org}).
While these surveys do not include a dedicated plan for spectroscopic 
observations of QSOs, they will provide the necessary candidate database 
of these extremely rare objects for spectroscopy. 

On the theoretical side, the calculations presented here could also be 
improved.  Better modeling could allow us to use a wider range of scales.
As large-scale, high-resolution simulations become increasingly feasible,
it will become possible to precisely measure both the bias of all halos
and the bias as a function of halo history.
The scatter between these various relationships can also be measured in
self-consistent numerical simulations of merger-driven black hole growth.
\citet{Hop07} presented a recent analysis of the observed and simulated
scatter between black hole mass and various host (bulge) galaxy properties
($\sigma$, host galaxy mass, effective radius) at lower redshifts than we
are considering here.  They find that the intrinsic scatter is generally
lower than the observed (upper limit) and as small as $\sim 0.2\,$dex for
combinations that produce a fundamental plane for black holes.
However, they also note that the strong correlation of black hole mass with
bulge properties implies that the connection to halo mass is only indirect
and dominated by the evolution in the typical gas fractions.
Measurement of the evolution of these quantities in full cosmological
simulations \citep[e.g.][]{DiM08} could provide measurements of the expected
scatter at the very high redshifts where QSOs are observed to be so strongly
biased.

Regardless of these future prospects, the strength of the present constraint
on scatter in a monotonic relation between QSO luminosity and halo mass is
already surprising. As noted in the introduction, we expect scatter to
be present because we expect scatter in at least four relationships
that combine to determine the instantaneous luminosity of a QSO in a
halo of a given mass: the relationship between instantaneous luminosity
and peak luminosity; the relationship between peak luminosity and black
hole mass; the relationship between black hole mass and bulge properties; and
the relationship between bulge and halo mass. These expectations
arise from the observed scatter in these individual relations at lower
redshifts--intriguingly, in all cases the observed scatter only sets an
upper limit on the intrinsic scatter.

For the relation between black hole mass and bulge velocity dispersion known
as the  $\mbh - \sigma$ relation \citep{ferrarese00,gebhardt00},
\citet{tremaine02} showed that the intrinsic dispersion is no larger than
0.25--0.3 dex (or $0.58 - 0.70$ in our natural log notation).
\citet{ferrarese02} explored the relation between $\mbh - M_h$ by using
rotation curve data and argued that the dispersion between black hole and
halo mass could be even less than the $\mbh-\sigma$ dispersion, although
she notes that the uncertainty in the transformation between circular velocity
and halo mass is not well characterized.
Finally, \citet{Kol06} measured the distribution in Eddington ratio,
$L_{\rm bol}/L_{\rm Edd}$, for a sample of $z=0.3-4$ QSOs and find the
distribution is well described as log-normal with a peak at 0.25 and a
dispersion of 0.3 dex.
As this observed dispersion must account for the uncertainty in the mass
estimator and bolometric correction in addition to $L_{\rm bol}/L_{\rm Edd}$,
this measurement also sets an upper limit to the intrinsic dispersion in
$L_{\rm bol}/L_{\rm Edd}$.

In addition to the small scatter in $L - M_h$, the data also suggest that
the duty cycle is approximately unity. This may be less surprising, given
their extremely high luminosities and masses of their supermassive black
holes.
If QSOs are limited to accreting at no more than the Eddington rate, the
luminosities of $z>6$ QSOs imply that their central, supermassive black
holes already exceeded $\mbh \sim 10^9\msun$ when the Universe was less than
$1\,$Gyr old \citep{Fan01}. Subsequent near-infrared spectroscopy provides
further evidence that their central black holes are indeed this massive
\citep{barth03,jiang07}. To create such massive black holes at early times
requires $\fon \sim 1$, modulo the uncertainty in the initial seed mass.
Together, these strong constraints on the duty cycle and scatter in $L-M_h$
for high-redshift QSOs provide important new information on how
supermassive black holes and their host halos grew at early times.

\medskip

We thank Yue Shen and Cris Porciani for helpful comments on an earlier draft
of this paper.
The simulations used in this paper were analyzed at the National Energy
Research Scientific Computing Center.
We thank CCAPP at The Ohio State University and the participants in the quasar
mini-workshop for a stimulating meeting. JDC and MW also are grateful to
CCAPP for hosting them for an extended visit.
MW is supported by NASA.

\end{document}